\documentclass[manuscript]{aastex631}
\usepackage{graphicx}
\usepackage{amsmath}
\usepackage{multirow}

\renewcommand{\i}{\mathrm{i}}
\shorttitle{Reproducing Mercury instabilities}
\begin{document}

\title{Simple physics and integrators accurately reproduce Mercury instability statistics}

\correspondingauthor{Dorian S. Abbot}
\email{abbot@uchicago.edu}

\author{Dorian S. Abbot}
\affiliation{Department of the Geophysical Sciences \\
The University of Chicago \\
Chicago, IL 60637 USA}

\author{David M. Hernandez}
\affiliation{Harvard--Smithsonian Center for Astrophysics \\
Cambridge, MA 02138, USA}
\affiliation{Department of Astronomy \\
Yale University \\
New Haven, CT 06511 USA}

\author{Sam Hadden}
\affiliation{Canadian Institute for Theoretical Astrophysics \\
University of Toronto \\
Toronto, ON M5S 3H8, Canada}

\author{Robert J. Webber}
\affiliation{Department of Computing \& Mathematical Sciences \\
California Institute of Technology \\
Pasadena, California 91125}

\author{Georgios P. Afentakis}
\affiliation{Department of Physics \\
The University of Chicago \\
Chicago, IL 60637 USA}

\author{Jonathan Weare}
\affiliation{Courant Institute of Mathematical Sciences \\ 
New York University \\ 
New York, NY 10012 USA}

%%%%%%%%%%%%%%%%%%%%%%%%%%%%%%%%%%%%%%%%%%%%%%%%%%%%%%%%%%%%%%%%%%%%
\begin{abstract}
The long-term stability of the Solar System is an issue of significant scientific and philosophical interest. The mechanism leading to instability is Mercury's eccentricity being pumped up 
so high that Mercury either collides with Venus or is scattered into the Sun. Previously, only three five-billion-year $N$-body ensembles of the Solar System with thousands of simulations have been run to assess long-term stability. We generate two additional ensembles, each with 2750 members, and make them publicly available at \texttt{https://archive.org/details/@dorianabbot}. We find that accurate Mercury instability statistics can be obtained 
by (1) including only the Sun and the 8 planets, (2) using a simple Wisdom-Holman scheme without correctors, (3) using a basic representation of general relativity, and (4) using a time step of 3.16 days. By combining our Solar System ensembles with previous ensembles we form a 9,601-member ensemble of ensembles. In this ensemble of ensembles, the logarithm of the frequency of a Mercury instability event increases linearly with time between 1.3 and 5 Gyr,
suggesting that a single mechanism is responsible for Mercury instabilities in this time range and that this mechanism becomes more active as time progresses. Our work provides a robust estimate of Mercury instability statistics over the next five billion years, outlines methodologies that may be useful for exoplanet system investigations, and provides two large ensembles of publicly available Solar System integrations that can serve as testbeds for theoretical ideas as well as training sets for artificial intelligence schemes.
\end{abstract}

%%%%%%%%%%%%%%%%%%%%%%%%%%%%%%%%%%%%%%%%%%%%%%%%%%%%%%%%%%%%%%%%%%%%
\section{Introduction} \label{sec:intro}

The long-term stability of the Solar System is important for our fundamental understanding of Solar System dynamics, for contextualizing the Solar System in comparison to exoplanet systems, and for evaluating the validity of the Copernican Principle. The most likely identified route to Solar System instability is Mercury's eccentricity being pumped up to a high enough value that it has a 
close encounter with Venus \citep{laskar1994large,laskar2009existence}, either colliding with Venus or being scattered into the Sun \citep{zeebe2015highly}. Mercury instability events are ultimately due to resonances of the secular system's modes \citep{LithwickWu2011secular_chaos,Boue2012,lithwick2014secular,batygin2015chaotic,MogaveroLaskar2021,mogavero2022origin}.

The probability that Mercury's orbit becomes unstable has been estimated by three previous ensembles of $N$-body simulations (first by \citet{laskar2009existence} and subsequently by \citet{zeebe2015highly,abbot2021rare}) with thousands of members integrated for five billion years using symplectic integrators \citep{hair06}. \citet{brown2020repository} ran a 96-member ensemble that resulted in no Mercury instability events, \citet{brown2022long} ran a multi-thousand-member ensemble including the effects of perturbations by passing stars, and there have been a few other very small ensembles \citep[e.g.,][]{ito2002long,mikkola2022overlong}. Each of the large Solar System ensembles was run with slightly different physical assumptions, numerical schemes, and time steps (Table~\ref{tab:ensembles}). We will show in section~\ref{sec:ensembles} that the Mercury instability statistics of the \citet{laskar2009existence} and \citet{zeebe2015highly} ensembles do not differ statistically, and that the Mercury instability statistic of the \citet{abbot2021rare} ensemble is significantly larger in the range of 4-4.75 billion years.

\begin{table}[]
\centering
\begin{tabular}{ p{0.15\textwidth}p{0.2\textwidth}p{0.2\textwidth}p{0.15\textwidth}p{0.15\textwidth}p{0.1\textwidth}}
\textbf{Ensemble} & \textbf{Bodies Included} & \textbf{Extra} & \textbf{General Relativity} & \textbf{Integrator} & \textbf{Time Step} \\
\hline
\citet{laskar2009existence} & Sun, 8 planets, Pluto, Earth's moon & Earth-moon tidal dissipation, Solar quadrupole moment & \citet{saha1994long} & \citet{laskar2001high} & 9.13 days\\
\citet{zeebe2015highly} & Sun, 8 planets, Pluto & & \citet{saha1994long} & \citet{rauch2002hnbody} & 4 days\\
\citet{abbot2021rare} & Sun, 8 planets & & \citet{nobili1986simulation} & \citet{wisdom1996symplectic} & 8.06 days \\
\end{tabular}
\caption{Comparison of the physical assumptions and numerical schemes used in the three previous large ensembles of Solar System integrations for five billion years. \citet{laskar2009existence} reduce their timestep for $e>0.4$ but do not say to what value. Additionally, because of the scheme they use, their time step is effectively 0.577 times the stated value \citep{hernandez2022stepsize}.
\citet{zeebe2015highly} reduces his time step to 1 day for $0.55<e<0.70$ and 0.25 days for $0.70<e<0.80$.}
\label{tab:ensembles}
\end{table}

\citet{hernandez2022stepsize} showed that all existing ensembles are subject to numerical chaos due to time stepping, and this artificial instability can be mitigated by using smaller time steps. They postulated that the resulting artificial diffusion could affect phase space statistics such as the Mercury instability probability.  To avoid numerical chaos, they suggested a conservative time step that is $\frac{1}{16}$ of the effective period at pericenter of the innermost planet, following \cite{wisdom2015resolving},
\begin{equation}
    dt = \frac{\pi}{8}\sqrt{\frac{(1-e)^3}{1+e} \frac{a^3}{GM}},
    \label{eq:timestep}
\end{equation}
where $e$ is eccentricity, $a$ is semi-major axis, $M$ is the self-gravitating mass, and $G$ is the gravitational constant. Close encounters between Mercury and Venus occur when Mercury's eccentricity is about 0.85, which corresponds to a time step of 0.23 days according to Eq.~(\ref{eq:timestep}). 

In this paper, we present two new large ensembles (2750 members) of the Solar System for the next five billion years. Except for the integrator (see section~\ref{sec:model}), both use a similar numerical set up as \citet{brown2020repository,abbot2021rare,brown2022long}, but the first (\texttt{Fix dt}) has a fixed time step of $\sqrt{10}\approx 3.16$~days and the second (\texttt{Var dt}) uses a time step that shrinks as Mercury's eccentricity increases so as never to violate the \citet{wisdom2015resolving,hernandez2022stepsize} criterion (Eq.~\eqref{eq:timestep}). The Mercury instability statistics of both \texttt{Fix dt} and \texttt{Var dt} ensembles are consistent with those of \citet{laskar2009existence} and \citet{zeebe2015highly} (section~\ref{sec:ensembles}), suggesting that the Mercury instability statistics are robust to fairly large differences in physical assumptions and numerical schemes (Table~\ref{tab:ensembles}). In section~\ref{sec:convergence} we perform convergence tests related to the Mercury instability problem and find that a time step of $\approx$9 days is sufficient to ensure convergence for these tests. We discuss our results in section~\ref{sec:discussion} and conclude in section~\ref{sec:conclusions}.

%%%%%%%%%%%%%%%%%%%%%%%%%%%%%%%%%%%%%%%%%%%%%%%%%%%%%%%%%%%%%%%%%%%%
\section{Model} \label{sec:model}

We perform all simulations using the REBOUND $N$-body code's \citep{rein2012rebound} \texttt{WHFAST} integration scheme \citep{rein2015whfast}, which is a Wisdom-Holman scheme (WH) \citep{WH1991}. We do not use REBOUND's \texttt{WHCKL} scheme, as \citet{brown2020repository,abbot2021rare,brown2022long} did, based on results from \cite{Hetal2020,hernandez2022stepsize}, suggesting there is no clear advantage to a higher order scheme for this problem.  We emphasize we have not used symplectic correctors \citep{wisdom1996symplectic}.  An integration improved by correctors indicates numerical instability is not present \citep{wisdom2015resolving,hernandez2022stepsize}, but correctors do not affect the existence of numerical instability appreciably. As in \citet{abbot2021rare}, we define a Mercury instability event as occurring when Mercury passes within 0.01 au of Venus and stop the simulations at that point. We do not attempt to resolve the close encounter of Mercury and Venus or subsequent behavior.

Our integrations contain all 8 Solar System planets (except for the simulations with massless Mercuries in section~\ref{sec:convergence}) and include an approximation of general relativity with a modified position-dependent potential \citep{nobili1986simulation}, which is implemented as the \texttt{gr\_potential} scheme in REBOUNDx \citep{Tamayo2020reboundx}. We initialize the simulations with Solar System conditions on February 10, 2018 from the NASA Horizons database using standard REBOUND functionality. We show that roundoff error remains small throughout our 5 billion year integrations in the Appendix.

We produce two ensembles of simulations (\texttt{Var dt} and \texttt{Fix dt}) that both contain 2750 members. In \texttt{Var dt} we use the heartbeat functionality in REBOUND to halve the time step if the \citet{wisdom2015resolving,hernandez2022stepsize} criterion (Eq.~\eqref{eq:timestep}) is about to be violated. Specifically, Mercury's heliocentric distance, $r_M$, is monitored and the time step, $dt$, is halved whenever $r_M < a_M(1-e_\mathrm{crit}(dt))$ where $a_M\approx 0.387$ AU is Mercury's semi-major axis, which we approximate as fixed, and  $e_\mathrm{crit}(dt)$ is the critical eccentricity satisfying Equation \eqref{eq:timestep} for the current time step. We generate the \texttt{Var dt} ensemble by adding Gaussian perturbations to Mercury's $x$-position with a 1~cm scale. Because the Gaussian perturbations are so small and because of limited numerical precision, 510 of the simulations are identical to another simulation. Since the repeated simulations are unbiased with respect to Mercury instability events, we include them in our statistics. In the \texttt{Fix dt} ensemble we use a fixed time step of $\sqrt{10}\approx 3.16$~days. We initialize the \texttt{Fix dt} ensemble on a uniform grid of perturbations to Mercury's $x$-position, each separated by 10~cm. This avoids the duplicated simulation problem encountered with our initialization of the \texttt{Var dt} ensemble.

Our \texttt{Var dt} and \texttt{Fix dt} ensembles are publicly available at \texttt{https://archive.org/details/@dorianabbot}. We save the model state every 10,000 years throughout the simulations using the REBOUND simulation archive feature \citep{rein2017new}, resulting in a total of 5~TB of data.

%%%%%%%%%%%%%%%%%%%%%%%%%%%%%%%%%%%%%%%%%%%%%%%%%%%%%%%%%%%%%%%%%%%%
\section{Solar System Ensembles} \label{sec:ensembles}

\begin{figure}[ht!]
\centering
\includegraphics[width=0.5\linewidth]{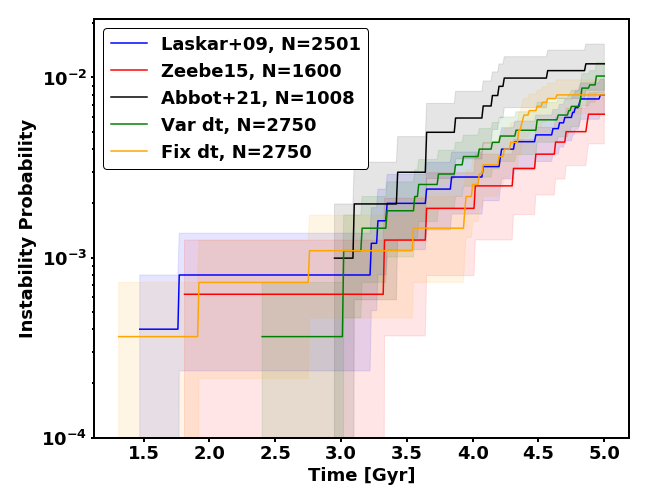}
\caption{Mercury instability probability as a function of time for the \citet{laskar2009existence} ensemble (2501 members, blue), the \citet{zeebe2015highly} ensemble (1600 members, red), the \citet{abbot2021rare} ensemble (1008 members, black), and the \texttt{Var dt} (2750 members, green) and \texttt{Fix dt} (2750 members, orange) ensembles from this paper. The shading represents 1$\sigma$ error bars using the fact that the observed instability events are binomial random variables.}
\label{fig:ensembles}
\end{figure}

Fig.~\ref{fig:ensembles} contains Mercury instability statistics as a function of time for the three previous large $N$-body Solar System ensembles  \citep{laskar2009existence,zeebe2015highly,abbot2021rare} and for our two new ones, one with a variable time step (\texttt{Var dt}) and one with a fixed time step (\texttt{Fix dt}). The author of \citet{zeebe2015highly} shared his data with us and we obtained the instability times from \citet{laskar2009existence} using a plot digitizer. We estimate 1$\sigma$ error bars for all ensembles using the fact that the observed instability events are binomial random variables \citep{abbot2021rare}.

\begin{figure}[ht!]
\centering
\includegraphics[width=0.5\linewidth]{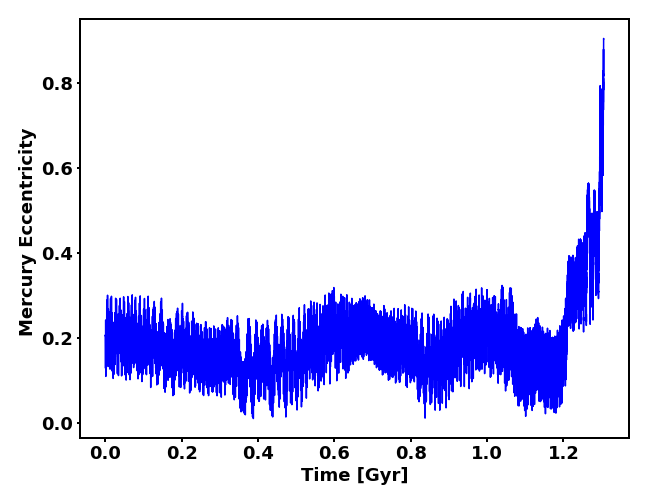}
\caption{Mercury's eccentricity as a function of time for the simulation from the \texttt{Fix dt} ensemble in which Mercury's orbit becomes unstable the fastest.}
\label{fig:fast_unstable}
\end{figure}

Despite differences in physical and numerical assumptions, the ensembles produce broadly similar results. All ensembles show a roughly exponential increase in the number of events after 3 billion years, from a probability of $\sim 10^{-3}$ at 3 billion years to $\sim 10^{-2}$ at 5 billion years. Before 3 billion years the ensembles have 2 or fewer Mercury instability events, which is too few to produce a discernible pattern. Our \texttt{Var dt} and \texttt{Fix dt} ensembles produce Mercury instability events before 3 billion years, like those of \citet{laskar2009existence} 
and
\citet{zeebe2015highly}. Our \texttt{Fix dt} ensemble produces the earliest Mercury instability event of any of the ensembles (Fig.~\ref{fig:fast_unstable}). The main reason the first Mercury instability event from our ensemble from \citet{abbot2021rare} occurs at 2.95 billion years is likely that the ensemble only had 1008 members. We were able to produce an ensemble with earlier Mercury instability events in 
\citet{abbot2021rare} using a rare event simulation algorithm.

It is tempting to try to interpret the differences among the ensembles. Before doing so, it is important to determine whether any of the differences are statistically significant. To do this we need a statistical test to compare the rate of Mercury instability events in two populations (two different ensembles).
Here, we use the statistical test that is recommended in \citet{d1988appropriateness}.
Assume that the true proportion of Mercury instability events in population 1 is $p_1$ and the proportion in population 2 is $p_2$. The null hypothesis is that $p_1=p_2$, or $H_0: p_1=p_2$. The alternative hypothesis is $p_1 \neq p_2$, or $H_a: p_1 \neq p_2$. Let us take a random draw of $n_1$ members from population 1 and observe $x_1$ events for a sampled rate of $\hat{p}_1=\frac{x_1}{n_1}$. Let us take a random draw of $n_2$ members from population 2 and observe $x_2$ events for a sampled rate of $\hat{p}_2=\frac{x_2}{n_2}$. Now define the overall sample proportion as
\begin{equation}
    \hat{p}=\frac{x_1+x_2}{n_1+n_2}=\frac{n_1\hat{p}_1+n_2\hat{p}_2}{n_1+n_2}.
    \label{eq:p-hat}
\end{equation}
We form the $z$ statistic
\begin{equation}
    z=\frac{\hat{p}_1-\hat{p}_2}{\hat{\sigma}},
    \label{eq:z}
\end{equation}
where our estimate of the standard error is
\begin{equation}
\hat{\sigma}=\sqrt{\hat{p}(1-\hat{p})\bigl(\frac{1}{n_1}+\frac{1}{n_2}\bigr)}.
\label{eq:sigma-hat}
\end{equation}
Since we have no reason to believe either population is larger {\it a priori}, we need to do a two-tailed test using this z statistic.

\begin{figure}[ht!]
\centering
\includegraphics[width=0.5\linewidth]{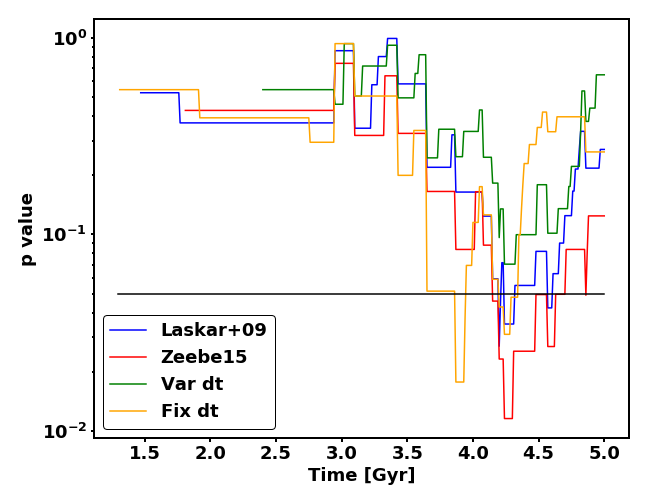}
\caption{p-value testing the Null Hypothesis that the Mercury instability event rate of the plotted ensemble is equal to the rate of the \citet{abbot2021rare} ensemble using the test described in Eq.~(\ref{eq:p-hat}-\ref{eq:z}). The black line represents p=0.05. The ensembles tested are the \citet{laskar2009existence} ensemble (blue), the \citet{zeebe2015highly} ensemble (red), the \texttt{Var dt} ensemble (green) and \texttt{Fix dt} ensemble (orange).}
\label{fig:stats}
\end{figure}

When we perform this test on each combination of the ensembles of \citet{laskar2009existence}, \citet{zeebe2015highly}, \texttt{Var dt}, and \texttt{Fix dt}, we find that they are statistically consistent over the entire 5 billion year period tested. This is not surprising given the overlap of the 1$\sigma$ error bars in Fig.~\ref{fig:ensembles}. Fig.~\ref{fig:stats} shows the results of the statistical test comparing the \citet{abbot2021rare} ensemble to the other four ensembles. The \citet{abbot2021rare} ensemble is statistically different from all of them ($p=0.05$) except \texttt{Var dt} at some point between 3.75--4.75 billion years. This is consistent with the claim that the time step in the \citet{abbot2021rare} ensemble is too large \citep{hernandez2022stepsize}.

\begin{figure}[ht!]
\centering
\includegraphics[width=0.5\linewidth]{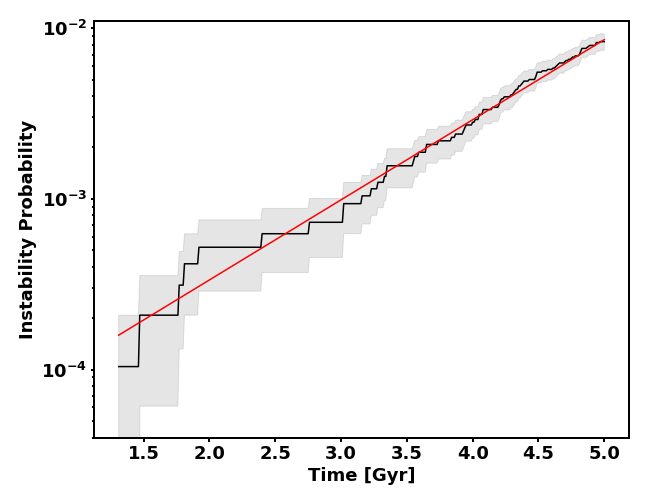}
\caption{Mercury instability probability as a function of time for the combined ensemble of ensembles including \citet{laskar2009existence}, \citet{zeebe2015highly}, \texttt{Var dt}, and \texttt{Fix dt} (9601 members). The shading represents 1$\sigma$ error bars using the fact that the observed instability events are binomial random variables. The red line shows the line of best fit for the logarithm of the probability of a Mercury instability event as a function of time in billions of years (Eq.~\ref{eq:linfit}).}
\label{fig:combined_ensemble}
\end{figure}

Since we cannot distinguish among the \citet{laskar2009existence}, \citet{zeebe2015highly}, \texttt{Var dt}, and \texttt{Fix dt} ensembles statistically, we can combine them to form one large ensemble of ensembles with 9601 members (Fig.~\ref{fig:combined_ensemble}). The logarithm of the probability that a Mercury instability event is approximately linear in time between 1.3 and 5 Gyr and can be fit to the line
\begin{equation}
    \log_{10} P = A + B T,
    \label{eq:linfit}
\end{equation}
where $A=-0.413 \pm 0.013$ and $B = 0.469 \pm 0.004$. This indicates that the probability of a Mercury instability event grows exponentially by almost an order of magnitude for every 2 billion years of time elapsed. If Eq.~(\ref{eq:linfit}) holds true for times earlier than 1.3 Gyr, it suggests that the probability of a Mercury instability event is about $6.6\times10^{-5}$ in the next 0.5 billion years and about $1.1\times10^{-4}$ in the next one billion years.

%%%%%%%%%%%%%%%%%%%%%%%%%%%%%%%%%%%%%%%%%%%%%%%%%%%%%%%%%%%%%%%%%%%%
\section{Convergence Tests} \label{sec:convergence}
%%%%%%%%%%%%%%%%%%%%%%%%%%%%%%%%%%%%%%%%%%%%%%%%%%%%%%%%%%%%%%%%%%%%

In this section we perform two convergence tests relevant to Mercury instability events using the Wisdom-Holman scheme as implemented in the REBOUND $N$-body code.  In the first test, we run simulations of the Solar System with one thousand massless versions of Mercury and determine the time step necessary to get numerically stable results. In the second test, we do the same thing using one thousand massless versions of Mercury at different semi-major axes. In both tests, we find that a time step of about 9 days is small enough to get converged ensemble statistics. This is much larger than the 0.23 day criterion predicted by \citet{wisdom2015resolving,hernandez2022stepsize} (Eq.~\eqref{eq:timestep}). 

\begin{figure}[ht!]
    \centering
    \includegraphics[width=0.5\linewidth]{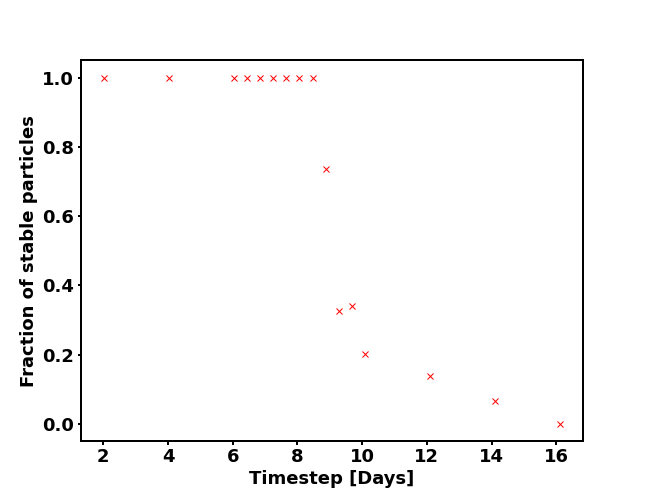}
    \caption{Fraction of 1000 massless ghost Mercuries that survive for 1 million years in a Solar System integration as a function of time step (red x's).}
    \label{fig:RealMercury}
\end{figure}

In our first convergence test, we run simulations of the Solar System with 1000 massless ghost Mercuries that are exact copies of Mercury except that they have zero mass and are initialized with true anomalies linearly spaced from 0 to $2\pi$. We perform 1-million-year simulations and study the fraction of ghost Mercuries whose eccentricity never exceeds 0.8. This fraction is 1 for a sufficiently small time step. We find that the model has converged at a time step of 9.1 days (Fig.~\ref{fig:RealMercury}). We extended some of our simulations for 10 million years and found similar results.

In our second convergence test, we run simulations of a modified Solar System with the Sun and all of the planets except Mercury. We then add 1000 massless ghost Mercuries at a value of the semi-major axis ($a$) that we vary among the different ensembles. All ghost Mercuries start with an eccentricity of 0.2 and are initialized with true anomalies linearly spaced from 0 to $2\pi$. For each value of $a$ we perform 1-million-year simulations and study the fraction of ghost Mercuries whose eccentricity never exceeds 0.8. For $a<0.6$ this fraction is 1 for a sufficiently small time step. For $a=0.6$ and $a=0.7$, we remove Venus and again this fraction is 1 for a sufficiently small time step. We consider an ensemble converged at the largest time step for which the fraction of stable ghost Mercuries is 1. 

\begin{figure}[ht!]
    \centering
    \includegraphics[width=0.5\linewidth]{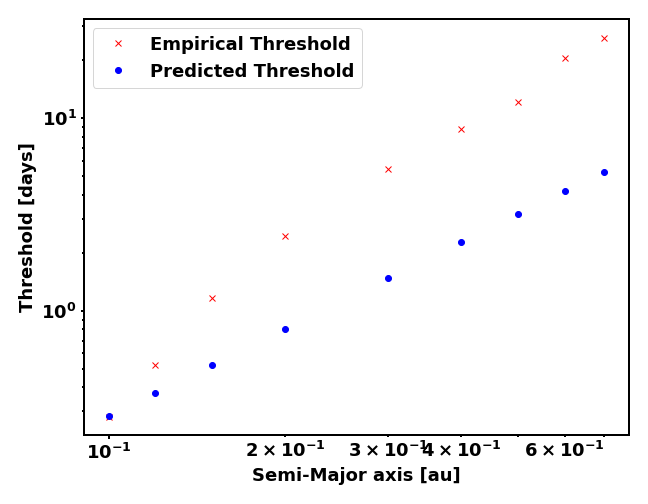}
    \caption{Our empirical time step threshold for convergence for the simulations with 1000 massless ghost Mercuries at various fixed semi-major axes (red x's) as well as the predicted time step threshold assuming an eccentricity of 0.4 (Eq.~(\ref{eq:timestep}), blue circles).}
    \label{fig:ThersholdVarySemiMajor}
\end{figure}

In Figure~\ref{fig:ThersholdVarySemiMajor} we plot the empirical time step threshold for convergence as a function of the semi-major axis of the ghost Mercuries. For $a=0.1$, we find that Eq.~(\ref{eq:timestep}) is a good predictor of empirical time step threshold if we use an eccentricity of 0.4, under the assumption that most ghost Mercuries that reach this eccentricity will become unstable quickly. As $a$ increases, the empirical time step threshold grows relative to Eq.~(\ref{eq:timestep}) such that it is 3--4 times larger than the predicted time step threshold for $a \geq 0.3$. For these values of $a$ we find a slope in the loglog plot of empirical time step threshold vs. semi-major axis with a 95\% confidence interval of 1.5--2.3, which is consistent with the prediction of $a^{1.5}$ in Eq.~(\ref{eq:timestep}), although possibly higher. Finally, for $a=0.4$, which is the closest semi-major axis we tested to Mercury's true value, we find an empirical time step threshold of 8.8~days.

%%%%%%%%%%%%%%%%%%%%%%%%%%%%%%%%%%%%%%%%%%%%%%%%%%%%%%%%%%%%%%%%%%%%
\section{Discussion} \label{sec:discussion}

It might seem surprising that our \texttt{Fix dt} and \texttt{Var dt} ensembles produce Mercury instability statistics indistinguishable from those of \citet{laskar2009existence} 
and
\citet{zeebe2015highly} (Fig.~\ref{fig:ensembles}), given the seemingly large differences in physical assumptions and numerical schemes (Table~\ref{tab:ensembles}). This suggests that physical factors such as Earth's Moon, Pluto, Earth-Moon tidal dissipation, and the Solar quadrupole moment do not significantly affect Mercury instability statistics. Moreover, although including general relativity has an enormous effect on Mercury instability statistics \citep{laskar2009existence}, the detailed specification of general relativity does not have a large impact. Finally, it is reasonable that the order of the Wisdom-Holman scheme does not impact Mercury instability statistics, since \citet{hernandez2022stepsize} showed that higher order Wisdom-Holman methods do not impact the onset of numerical chaos.

This work raises the question of why the \citet{wisdom2015resolving,hernandez2022stepsize} criterion (Eq.~(\ref{eq:timestep})) predicts a time step more than an order of magnitude smaller than what we find is necessary for convergence of ensemble Mercury instability statistics. A major part of the explanation is likely that it is not necessary to correctly resolve Mercury's orbit all the way until a collision or scattering with Venus in order to correctly infer statistics in an ensemble. Most ensemble members in which Mercury's eccentricity exceeds 0.4 for any significant period of time experience a Mercury instability event eventually \citep{laskar2009existence,zeebe2015highly,abbot2021rare}. Eq.~(\ref{eq:timestep}) predicts a time step of 2.2 days for an eccentricity of 0.4, which is almost ten times larger than the 0.23 days it predicts for an eccentricity of 0.85. 2.2 days is similar to the timestep of 3.16 days that we used in the \texttt{Fix dt} ensemble that produced statistically identital results to the \texttt{Var dt} ensemble. It is interesting to note that we found a criterion of about 9 days for convergence in section~\ref{sec:convergence}. This suggests that it might be possible to get converged results with a somewhat larger fixed time step. We should also note, however, that failure in the convergence tests we performed in section~\ref{sec:convergence} was defined by numerical noise dominating the system on a short time scale. It would not be surprising if a more stringent constraint on the time step is necessary to reproduce more nuanced physical variables in 5~Gyr integrations. Finally, the fact that Mercury instability statistics converge at a time step more than an order of magnitude larger than the conservative \citet{wisdom2015resolving,hernandez2022stepsize} criterion (Eq.~(\ref{eq:timestep})) does not imply that the same is true for all other statistics of other functions of phase space.

If we consider a single one of the \citet{laskar2009existence}, \citet{zeebe2015highly}, \texttt{Var dt}, and \texttt{Fix dt} ensembles, it might appear that Mercury instability events have different characteristics before and after 3~Gyr, potentially pointing to different instability mechanisms at different time scales. However, we find that a single line (Eq.~\ref{eq:linfit}) fits well the logarithm of the probability of Mercury instability events as a function of time for the ensemble of these ensembles between 1.3 and 5 Gyr. This suggests that a single mechanism is responsible for Mercury instability events between 1.3 and 5 Gyr, and that this mechanism becomes more pronounced with time.

Using the line of best fit (Eq.~\eqref{eq:linfit}), we can estimate the probability of a Mercury instability event occurring in the next two billion years to be $3.3\times10^{-4}$. In \citet{abbot2021rare}, we estimated the probability of a Mercury instability event occurring in the next two billion years using QDMC, a rare event sampling algorithm. Our estimate here is consistent with our estimate in \citet{abbot2021rare} when we used a target time for QDMC of 2.8 billion years. Our estimate here is about an order of magnitude higher than our estimate in \citet{abbot2021rare} when we used a target time for QDMC of 2.4 billion years, but we only observed one event in the first two billion years in that ensemble, so we would expect high variance in that estimate.

The large data set of Solar System futures we have generated could be useful for further data analysis and training artificial intelligence schemes. For example, the data set could be mined to identify system characteristics that predict Mercury instability events and the time horizon for meaningful predictions. Techniques such as this have already been used in ocean-atmosphere dynamics \citep{tantet2015early,chattopadhyay2020analog,wang2020extended,Finkel2020paths,Finkel2021learning,Finkel2022data,Miloshevich2022probabilistic,jacques2022data} and chemistry \citep{ma2005nncommittor,Thiede2019galerkin}.

%%%%%%%%%%%%%%%%%%%%%%%%%%%%%%%%%%%%%%%%%%%%%%%%%%%%%%%%%%%%%%%%%%%%
\section{Conclusions} \label{sec:conclusions}

The main conclusions of this paper are:

\begin{enumerate}
    \item We produce two new publicly available 2750-member ensembles for the future evolution of the planets in the Solar System for 5 billion years. When we combine these ensembles with previous ensembles we find that the logarithm of the probability of Mercury instability events is linear in time between 1.3 and 5 billion years. This suggests that a single mechanism which becomes more pronounced with time is responsible for Mercury instability events in this time frame. 
    \item Accurate Mercury instability statistics over the next 5 billion years can be obtained using a simple Wisdom-Holman scheme without correctors and a representation of general relativity, applied to the Sun and the 8 planets. 
    \item A time step of 3.16 days is sufficient for numerical stability and converged ensemble Mercury instability statistics.
    \item One of our ensembles produced a Mercury instability event 1.3 billion years in the future, which is the soonest ever observed with an $N$-body simulation.
\end{enumerate}

%%%%%%%%%%%%%%%%%%%%%%%%%%%%%%%%%%%%%%%%%%%%%%%%%%%%%%%%%%%%%%%%%%%%
\section*{acknowledgments}
We thank Hanno Rein for helpful discussions about numerical schemes. We thank Richard Zeebe for sharing his simulation data with us. We thank an anonymous reviewer, Richard Zeebe, and Jacques Laskar for comments on a draft of this manuscript. This work was completed with resources provided by the University of Chicago Research Computing Center. This work was supported by the NASA Astrobiology Program grant No. 80NSSC18K0829 and benefited from participation in the NASA Nexus for Exoplanet Systems Science research coordination network.

\section*{software}

This research made use of the open-source projects Jupyter \citep{Kluyver2016jupyter}, iPython \citep{PER-GRA:2007}, and matplotlib \citep{Hunter:2007}.

%%%%%%%%%%%%%%%%%%%%%%%%%%%%%%%%%%%%%%%%%%%%%%%%%%%%%%%%%%%%%%%%%%%%
\appendix
\section{Roundoff error small in our simulations}

\begin{figure}[ht!]
\centering
\includegraphics[width=0.5\linewidth]{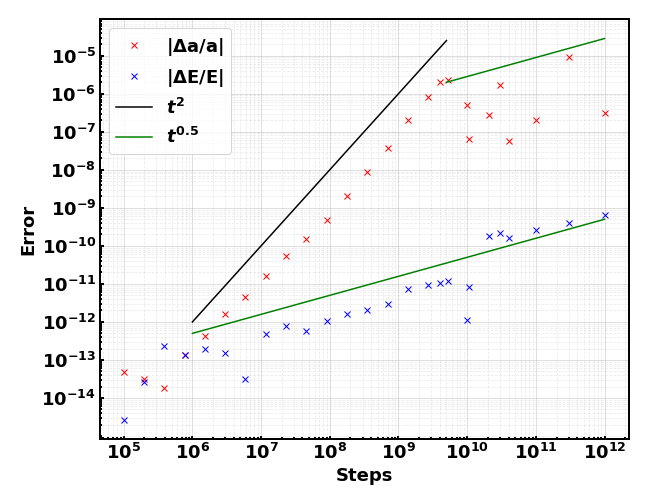}
\caption{Accumulation of fractional error in total energy ($E$) and Mercury's semi-major axis ($a$) when a single realization of the solar system is run forward and backward for a given number of total time steps. The time step is 3.16 days and REBOUND is configured and initialized as described in section~\ref{sec:model}. A 5~Gyr simulation corresponds to 5.8$\times$10$^{11}$ steps.}
\label{fig:ForwardBackward}
\end{figure}

Fig.~\ref{fig:ForwardBackward} shows the accumulated error in total energy and Mercury's semi-major axis when a single realization of the solar system is run forward and backward for a given number of total time steps using the REBOUND set up described in section~\ref{sec:model}.  In the absence of roundoff error the errors would be $0$. The maximum number of time steps we performed this test for is 10$^{12}$, which is larger than the 5.8$\times$10$^{11}$ steps necessary to run 5~Gyr with a 3.16 day time step. The energy error shows a simple $t^{0.5}$ scaling, indicating unbiased action-variable-like growth of roundoff error \citep{brouwer1937accumulation}. After initial unbiased behavior (see Fig. \ref{fig:bias}), the error in Mercury's semi-major axis grows like $t^2$ until 10$^{9.5}$ steps have passed, indicating biased angle-like roundoff error growth. After this the error grows like 
$t^{0.5}$, indicating unbiased action-variable-like growth of roundoff error. The error both in the energy and Mercury's semi-major axis remain relatively small after 10$^{12}$ steps, which suggests that roundoff error should not significantly affect our 5~Gyr integrations.

\section{General Relativity and biased roundoff error}

\begin{figure}[ht!]
\centering
\includegraphics[width=0.5\linewidth]{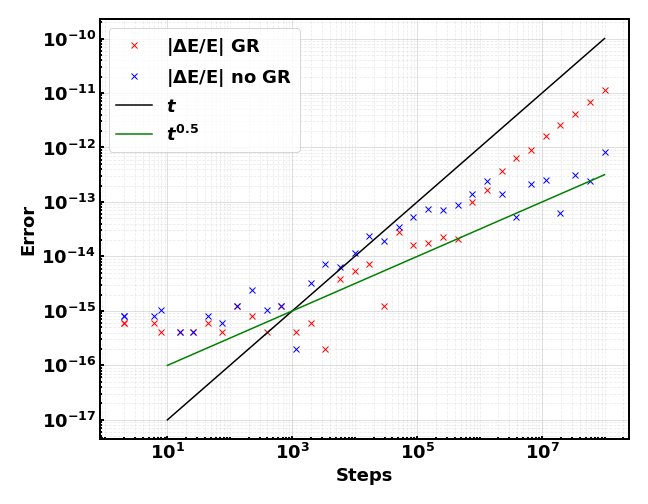}
\caption{Accumulation of fractional error in energy including GR (red) and not including GR (blue) as a function of number of time steps for a REBOUND simulation including only the Sun and Mercury that is run forwards only.}
\label{fig:bias}
\end{figure}

The integrator WHFast has been shown to be unbiased for specific problems, number of steps, and phase space variables \citep{rein2015whfast,rein2017new}.  However, it is possible that relativistic corrections affect this behavior.  Here we show how including a simple general relativistic potential can introduce bias in this integrator.  For this test, we use Gaussian units, and consider a system composed only of the Sun and Mercury.  The initial conditions are as follows:
\begin{quote}
\texttt{sim.add( m=1.00000597682, x=3.256101656448802E-03, y=-1.951205394420489E-04, z=-1.478264728548705E-04, vx=3.039963463108432E-06, vy=6.030576499910942E-06, vz=-7.992931269075703E-08)} \# Sun \\
\texttt{ sim.add( m=1.66051141e-7,   x=-1.927589645545195E-01,
y=2.588788361485397E-01,
z=3.900432597062033E-02,   vx=-2.811550184725887E-02,  vy=-1.586532995282261E-02,
vz = 1.282829413699522E-03) \# Mercury}.
\end{quote}
The GR correction from \cite{nobili1986simulation} is used, and we calculate energy error over time with time step $0.0543988406$ days.  This is not a forward--backward integration as in Fig. \ref{fig:ForwardBackward}.  We also compare against an integration without GR correction.  The error grows as $\sqrt{N}$, where $N$ is the number of time steps, when GR is neglected, but a bias appears, and the growth of error scales with $N$, when GR is included. We tested a larger timestep of $3.2$ days and find that bias appears after a similar number of timesteps as in Fig. \ref{fig:bias}.  The step number at which roundoff error becomes biased also agrees with the onset of bias in the semi-major axis of Mercury from our roundoff studies of the full Solar System in Fig. \ref{fig:ForwardBackward}.

%%%%%%%%%%%%%%%%%%%%%%%%%%%%%%%%%%%%%%%%%%%%%%%%%%%%%%%%%%%%%%%%%%%%
\bibliography{convergence}{}
\bibliographystyle{aasjournal}

\end{document}